\documentclass[fleqn,twoside]{article}%
\usepackage{amsmath}
\usepackage{amsfonts}
\usepackage{amssymb}
\usepackage{graphicx}%
\def\be{\begin{equation}}

\def\ee{\end{equation}}
\def\bes{\begin{equation}\begin{split}&}
\def\es{\end{split}}
\def\bi{\bibitem}

\begin{document}
\title{\boldmath On the equivalence between different canonical forms of F(R) theory of gravity.}
\author{Nayem Sk$^\dag$, Abhik Kumar Sanyal$^\ddag$}
\maketitle
\noindent
\begin{center}
\noindent
$^{\dag}$ Dept. of Physics, University of Kalyani, West Bengal, India - 741235.\\
\noindent
$^{\ddag}$ Dept. of Physics, Jangipur College, Murshidabad,
\noindent
West Bengal, India - 742213. \\
\end{center}
\footnotetext[1] {\noindent
Electronic address:\\
\noindent
$^{\dag}$nayemsk1981@gmail.com\\
$^{\ddag}$sanyal\_ ak@yahoo.com}
\begin{abstract}

\noindent
Classical equivalence between Jordan's and Einstein's frame counterparts of F(R) theory of gravity has recently been questioned, since the two produce different Noether symmetries, which couldn't be translated back and forth using transformation relations. Here we add the Hamiltonian constraint equation, which is essentially the time-time component of Einstein's equation, through a Lagrange multiplier to the existence condition for Noether symmetry to show that all the three different canonical structures of F(R) theory of gravity, including the one which follows from Lagrange multiplier technique, admit each and every available symmetry independently. This establishes classical equivalence amongst all the three.\\

\end{abstract}

\noindent
keywords:\\
F(R) theory; Lagrangian Multiplier Technique; Jordan's Frame, Einstein's Frame; Equivalence.
\maketitle

\section{Introduction}

In a recent publication \cite{1}, mathematical equivalence between Jordan's and Einstein's frame in connection with $F(R)$ theory of gravity described by the action

\be\label{1.1} S = \int  d^4 x\sqrt{-g}\left[\frac{1}{2\kappa} F(R)+ {\mathcal{L}_m}\right] + {1\over \kappa}\int_{\partial\mathcal{V}}\sqrt{h} ~d^3 x F'(R) K,\ee
has been questioned. In the above action \eqref{1.1}, $\kappa = 8\pi G$, $\mathcal{L}_m$ represents matter Lagrangian density, $h$ is the determinant of the induced metric, $K$ is the trace of the extrinsic curvature tensor $K_{ij}$, and prime denotes derivative with respect to the Ricci scalar $R$. The question arose because, the two produce different symmetries even in homogeneous and isotropic model, which can't be translated back and forth using the transformation relation

\be\label{1.2}\tilde \phi = \sqrt{3\over 2k} \ln \Phi.\ee
while it establishes mathematical equivalence between the two, by translating the classical field equations. To define the scalars $\Phi$ and $\tilde\phi$, we mention that, translation of action \eqref{1.1} to the Jordan's frame of reference is possible, simply under redefinition of $F'(R) = \Phi$ and $R = U_{,\Phi}$, while the same in Einstein's frame of reference is possible under conformal transformation, $\tilde{g}_{\mu\nu} = F'(R) g_{\mu\nu} = e^{2\omega} g_{\mu\nu}$, where the conformal factor $\omega$ is related to an effective scalar field $\tilde\phi$ by the relation $\omega = \sqrt{k\over 6}\tilde\phi$.\\

In fact, there exists three possible techniques towards canonical formulation of $F(R)$ theory of gravity. One is Lagrange multiplier technique, which is applicable with finite degrees of freedom, and the other two as mentioned, are the Jordan's and Einstein's frame counterparts of the above action \eqref{1.1}. Invoking Noether symmetry taking into account barotropic fluid, for which $ {\mathcal{L}_m} =  - \rho_{0} a^{-3w}$ (where $\rho_0$ is the presently available matter-density of the universe, and $w$ is the state parameter) in Robertson-Walker minisuperspace,

\be\label{1.3} ds^2 = -dt^2 + a(t)^2 \left[{dr^2\over 1-kr^2} + r^2 d\theta^2 + r^2 \sin^2 \theta d\phi^2\right],\ee
while Lagrange multiplier technique \cite{1, 2, 3, 4, 5, 6, 7, 8, 9, 10} having point Lagrangian in the form,

\bes\label{1.4} L(a,R,\dot{a},\dot{R}) = {1\over 2\kappa} \left[a^3(F-R F') -6a(\dot a^2-k) F' -6a^2 \dot a\dot R F''\right]- \rho_{0} a^{-3w}.\end{split}\ee
and the point Lagrangian corresponding to equivalent canonical action in Jordan's frame \cite{1, 11, 12, 13, 14, 15}

\bes\label{1.5} L(a,\Phi,\dot a,\dot \Phi) =\frac{1}{16\pi G} [- 6a^2 \dot a \dot \Phi - 6a \dot a^2\Phi+ 6\Phi a k- a^3U(\Phi)]
 -\rho_0 a^{-3w}\\& \mathrm{with},~U(\Phi) = R F'(R)- F(R);\;\;\;\Phi = F'(R)\end{split}\ee
yield only

\be\label{1.6} F(R) \propto R^{3\over 2};\;\;\;~~~ \mathrm{with},~\Sigma = {d\over dt} a\sqrt R,~~~~~~~~\mathrm{for} ~ k = 0, \pm 1,~\mathrm{in ~vacuum ~and ~pressureless ~dust},\ee
the point Lagrangian corresponding to equivalent canonical action in Einstein's frame \cite{1, 13, 14, 15, 16, 17, 18, 19, 20}

\bes\label{1.7} \tilde L_{\tilde t}[\tilde a,\tilde \phi,\dot{\tilde a},\dot{\tilde \phi}]=\frac{3}{\kappa}(-\tilde a \dot{\tilde a}^2
+k\tilde a)+\left({1\over 2}\dot{\tilde\phi}^2 - \tilde V(\tilde\phi)\right)\tilde a^3
- \rho_0\tilde a^{-3w}e^{-{\sqrt {\kappa\over 6}}(1-3w)\tilde \phi},\\&
\mathrm{where},~ \tilde V(\tilde\phi) = \frac {R F'(R)-F(R)}{2\kappa(F'(R))^2}.\end{split}\ee
with,

\be\begin{split}\label{1.8}& F(R) = F_0 R^2;~ ~~\mathrm{with},~ \Sigma= a^3\dot R,~~~~~~~~~~~~\mathrm{for} ~ k = 0, \pm 1 ~\mathrm{in ~vacuum ~and ~in ~radiation ~era},\\&
F(R) = {F_0\over R};~~~~~~ \mathrm{with},~ \Sigma = R{\sqrt a}\dot a,~~~~~~~~~ \mathrm{for}~k = 0, ~\mathrm{in ~vacuum},\\&
F(R) = F_{0}R^{\frac{7}{5}};~~~ \mathrm{with}, ~\Sigma=\sqrt a {d\over dt}(a R^{2\over 5}),~\mathrm{for}~k = 0, ~\mathrm{in ~vacuum}.\end{split}\ee
In the above, $\Sigma$ stands for the associated conserved current. It was not possible to translate the conserved currents back and forth using the transformation relations \eqref{1.2} and the following ones,

\be\label{1.9}\tilde a = a{\sqrt\Phi}, ~d\tilde t = {\sqrt\Phi} dt, \ee
which follow from the metric \eqref{1.3}. To understand the awesome situation, it is important to mention that the classical field equations in the two frames, do not match automatically. The scalar field equations in particular match, using the $(^0_0)$ equation of Einstein. It is also important to mention that since the $(^0_0)$ equation of Einstein doesn't contain second derivative, it simply is treated as a non-holonomic constraint of the theory of gravity, and when translated in terms of the phase-space variables, is found to be the Hamiltonian constraint $(H = 0)$ of the theory. In the absence of the constraint the classical equations of motion don't match and this was made clear in the recent manuscript \cite{1} in view of a toy model.\\

Since the classical field equations in the two frames match only due to the presence of the Hamiltonian constraint equation, which is the outcome of diffeomorphic invariance of the theory of gravity, therefore due to the presence of constraint, one can't expect that solutions of Noether equations satisfy the Hamilton constraint equation automatically. This has been proved by Wald and Zoupas \cite{21}. This means that Noether theorem is not on-shell for constrained system, unless the symmetry generator involves the constraint. We therefore propose that for such a constrained system, the symmetry generator obtained under infinitesimal transformation of the generalized coordinates, $q_i' = q_i + \epsilon \alpha_i(q_{i}, t)$ and time, $t'=t + \epsilon \chi(q_{i}, t)$, where, $\epsilon \rightarrow 0$, of the associated point Lagrangian, is

\be\label{1.10} \pounds_X L - \eta H = X L - \eta H = 0, ~\mathrm{where}, ~ \eta\ = {d\chi\over dt}.\ee
In the above, the Hamiltonian should be expressed in terms of configuration space variables, which essentially is the $(^0_0)$ equation of Einstein. Precisely, we mean to say that the ($^0_0$) equation of Einstein should be added to the existence condition ($\pounds_X L=X L = 0$) of Noether symmetry, through a Lagrange multiplier $\eta$. This may also be treated as a gauge.\\

In view of the above generator involving gauge, it is possible to generate identical symmetries for all the three point Lagrangians constructed using Lagrange multiplier technique, and from equivalent Jordan's and Einstein's frame actions of $F(R)$ theory of gravity. This we demonstrate in the following section. In the process, we show that all the lost symmetries from different frames are recovered. Thus, mathematical equivalence between all the three canonical structures of $F(R)$ theory of gravity is established. We conclude in section 3.

\section{Noether symmetries.}

As mentioned, we have not found the same symmetries in different canonical frames, and this is the reason for our botheration. It is generally argued that symmetries depend on the choice of configuration space variables. Indeed it's true, but only for trivial Noether symmetries, which appear due to the presence of a cyclic co-ordinate in the Lagrangian. However, when Noether symmetry analysis is performed to find a potential from an arbitrary action, it is always there, irrespective of the choice of configuration space variables. For example, if we impose Noether symmetry in view of the condition $\pounds_X L = 0$, where the vector field, $X = \sum_i\left[\alpha_i\frac{\partial}{\partial q_i}+\dot\alpha_i\frac{\partial}{\partial\dot q_i}\right]$, to find the form of potential in the toy model
\be L = \frac{1}{2}m(\dot x^2 +\dot y^2) -V(x,y).\ee
It indeed exists in the form, $V(x,y) = V_0e^{c(x^2+y^2)}$. One can find a new set of variables say $u(q_i)$ and $v_j(q_i)$, ($j = i-1$) under the change of variables,
\be\label{N.4}i_X du = 1,\;\;i_X dv_j = 0.\ee
Once the Lagrangian is expressed in terms of the new variables, $u$ turns out to be a cyclic co-ordinate. In this case, the new variables are $u=\tan^{-1}{y\over x},\;\;v =\sqrt{x^2+y^2}$. One can now observe that in the process we have gone over to the polar coordinate $v = r,\;u = \theta$, for which the potential is $V = V_0e^{c r^2}$ and hence $u = \theta$ becomes cyclic.\\

We therefore claim that equivalence between different techniques may only be established, if all the symmetries corresponding to the action \eqref{1.1} so far obtained following different techniques are available in every technique.

\subsection{Symmetries following Lagrange multiplier technique.}

The Hamiltonian in terms of configuration space variables (the ($^0_0$) equation of Einstein), constructed out of the point Lagrangian \eqref{1.4} in the Robertson-Walker metric \eqref{1.3}, takes the form

\bes\label{2.1}H = {1\over 2\kappa} \left(a^3(R F' - F) -6a(\dot a^2+k) F' -6a^2 \dot a\dot R F''\right)+\rho_{0} a^{-3w}.
\end{split}\ee
\noindent
Now, in view of the vector field $X$

\be\label{2.2} X = \alpha \frac{\partial }{\partial {a}}+\beta\frac{\partial }{\partial R}+\dot\alpha \frac{\partial }{\partial\dot {a}}+\dot\beta \frac{\partial }{\partial \dot{R}},\ee
relative to the Lagrangian (\ref{1.4}) in the tangent space $(a, R, \dot{a}, \dot{R})$,
our proposed existence condition for symmetry \eqref{1.10}, $\pounds_X L- \eta H = X L - \eta H = 0 $, leads to the following differential equation,
\bes\label{2.3}
\frac{\alpha} {2\kappa} \Big [( 3a^2 (F - R F')- 6  \dot a^2 F'-12  a\dot a \dot R F'' +6 k F'+6\omega\kappa\rho_0a^{-(3\omega+1)}\Big]\\&+
 \frac{\beta} { 2\kappa}\Big [(6ka - a^3 R - 6 a \dot a^2) F'' - 6 a^2\dot a \dot R F'''\Big ]+\frac{1}{2\kappa}\Big [\frac{\partial\alpha}{\partial a}\dot a+\frac{\partial\alpha}{\partial R}\dot R  \Big](-12a\dot a F'
 -6a^2 \dot R F'')\\&+\frac{1}{2\kappa}\Big [\frac{\partial\beta}{\partial a}\dot a+\frac{\partial\beta}{\partial R}\dot R \Big](-6 a^2\dot a F'')-{\eta\over 2\kappa} \left[a^3(R F' - F) -6a(\dot a^2+k) F' -6a^2 \dot a\dot R F'' + 2\kappa\rho_0a^{-3\omega}\right] =0.\end{split}\ee
Therefore equating coefficients of ${\dot a^2}$, ${\dot R}^2$, $\dot { a} \dot {R}$  and the rest to zero, following set of four Noether equations are found

\bes\label{2.4}
F' (\alpha + 2a \alpha ,_a)+ (\beta  + a \beta ,_{a})a F'' - \eta a F' = 0, ~~\;\; a^2F''\alpha' = 0, \\& \beta a^2F''' +(2\alpha   +  a \alpha ,_{a} + a \beta')a F''+ 2 F'a \alpha' - \eta a^2 F''= 0, \;\;\\&
\alpha [3 (F - R F')a^2 + 6kF'+6\omega\kappa\rho_0a^{-(3\omega+1)}] + \beta [6k-a^2 R] a F''- \eta [a^3(R F' - F)-6ak F'+2\kappa\rho_0a^{-3\omega}]= 0.\end{split}\end{equation}

\subsubsection{Available symmetries in pure vacuum era.}

\noindent
\textbf{Case-I, $k = \pm 1, 0$}. This case results in two sets of solutions.\\

\noindent
\emph{Solution-I}\\
\be\label{2.1a}\alpha = a ,\;\;\;\beta = -2 R,\;\;\; \eta = 1,\;\;\;F(R) = F_0 R^2, \;\;\;\; {\Sigma}  =  a^3\dot R.\ee

\noindent
\emph{Solution-II}.\\
\be\label{2.1b}\alpha = \frac{n}{a},\;\;\;\;\beta = -\frac{2n R}{a^2},\;\;\;\; \eta=0,\;\;\;F(R) = F_0 R^{\frac{3}{2}}, \;\;\;\; {\Sigma} =\frac{d}{dt} ({a\sqrt R}),\ee

\noindent
\textbf{Case-II, $k = 0$}. Choosing $k = 0$, a priori, here again we obtain a pair of solutions.\\

\noindent
\emph{Solution-I}
\be\label{2.1c}\alpha = 0 ,\;\;\;\;\beta = -\beta_{0}a^{-\frac{3}{2}}R^{4} ,\;\;\;\; \eta= -\beta_{0}a^{-\frac{3}{2}}R^{3},\;\;\;F(R) =\frac{ F_0}{R} , \;\;\;\; {\Sigma} = R\dot a\sqrt a.\ee

\noindent
\emph{Solution-II}.\\
\be\label{2.1d}\alpha =\frac{\alpha_{0}}{2} a^{-\frac{1}{2}} ,\;\;\;\;\beta =-5\alpha_{0} R a^{-\frac{3}{2}} ,\;\;\;\; \eta=\frac{5\alpha_{0}}{2} a^{-\frac{3}{2}},\;\;\;F(R) = F_{0}R^{\frac{7}{5}} , \;\;\;\; {\Sigma} = \sqrt a {d\over dt}(a R^{2\over 5}).\ee

\subsubsection{Available symmetries in radiation dominated era.}

In the pure radiation era, $\omega=\frac{1}{3}$, and for, $k = \pm 1, 0$, $\eta=1$, the following set of solutions exists\\
\be \label{2.1e}\alpha = a,\;\;\;\;\beta = -2 R,\;\;\;\;\;\;\;\; F(R) = F_0 R^2, \;\;\;\; {\Sigma} =  a^3\dot R.\ee

\subsubsection{Available symmetries in pressureless dust era}

In the matter dominated era (pressureless dust) $\omega=0$, and for, $k = \pm 1, 0$, $\eta=0$, following set of solutions exists\\
 \be \label{2.1f}\alpha = \frac{n}{a},\;\;\;\;\beta = -\frac{2n R}{a^2},\;\;\;\;\;\;\;\; F(R) = F_0 R^{\frac{3}{2}}, \;\;\;\; {\Sigma _0} = \frac{d}{dt} ({a\sqrt R}).\ee
In all the different subsections, cases and subsubsections above, we have denoted constants by the same symbols, viz., $\alpha_0, \beta_0, F_0$ and $n$. One can now observe that while following Lagrange multiplier technique no symmetry other than $F(R) \propto R^{3\over 2}$, carrying a conserved current $\Sigma = {d\over dt} a\sqrt R$ was found earlier, despite enormous attempts over decades \cite{2, 3, 4, 5, 6, 7, 8, 9, 10}, here we have been able to explore various symmetries, which includes $F(R) \propto R^2$, $F(R) \propto R^{-1}$ and $F(R) \propto R^{7\over 5}$. The fact that $F(R)$ theory \eqref{1.1} must admit these symmetries was signalled while trying the same with Einstein's frame action analogue of the theory. Also it's important to mention that, $F(R) \propto R^2$ is a general symmetry for traceless field (as in pure vacuum and radiation eras) due to scale invariance of the action \cite{22}, and may be found directly from the field equations \cite{1}. Earlier it was not also possible to find the Noether counterpart of this general symmetry. In the process, we have recovered all the lost Noether symmetries following Lagrange multiplier technique, using the proposed symmetry generator which involves the Hamiltonian constraint \eqref{1.10}.

\subsection{Symmetries available in Jordan's frame of reference.}

Having recovered all the lost symmetries following Lagrange multiplier technique, we now turn our attention in this subsection, to explore symmetries from analogous Jordan's frame of reference. The Hamiltonian in terms of configuration space variables (the ($^0_0$) equation of Einstein), constructed out of the point Lagrangian \eqref{1.5} corresponding to the action \eqref{1.1} in Robertson-Walker metric \eqref{1.3} is given by \cite{1},
\be\label{2.5} H(a,\dot a,\Phi,\dot \Phi) = \frac{1}{2\kappa} [- 6a^2 \dot a \dot \Phi - 6a \dot a^2\Phi-6\Phi a k+ a^3U(\Phi)] +\rho_{0} a^{-3w}.\ee
Here again in view of the vector field $X$

\be\label{2.6} X = \alpha \frac{\partial }{\partial {a}}+\beta\frac{\partial }{\partial \Phi}+\dot\alpha \frac{\partial }{\partial\dot {a}}+\dot\beta \frac{\partial }{\partial \dot{\Phi}},\ee
relative to the Lagrangian $L$ (\ref{1.5}) in the tangent space $(a, \Phi, \dot{a}, \dot{\Phi})$, the existence condition for symmetry, $\pounds_X L -\eta H = X L -\eta H = 0 $ \eqref{1.10}, leads to the following  differential equation
\bes\label{2.7}  {\alpha\over 2\kappa} [- 12a \dot a \dot \Phi - 6 \dot a^2\Phi+ 6\Phi k-3a^2U(\Phi)+6\kappa\omega\rho_0a^{-(3\omega+1)}]+{\beta\over 2\kappa}[-6a \dot a^2+ 6 a k- a^3U_{,\Phi}]\\&+\frac{1}{2\kappa}(\alpha_{,a}\dot a+\alpha_{,\Phi}\dot \Phi)[-12a\dot{a}\Phi-6a^2\dot\Phi]+\frac{1}{2\kappa}(\beta_{,a}\dot a+\beta_{,\Phi}\dot \Phi)[-6a^2\dot{a}]\\&-\frac{\eta}{2\kappa}\left( \frac{1}{2\kappa} [- 6a^2 \dot a \dot \Phi - 6a \dot a^2\Phi- 6\Phi a k+ a^3U(\Phi)] + 2\kappa\rho_0a^{-3\omega} \right)=0.
\end{split}\ee

\noindent
Therefore, equating coefficients of ${\dot a^2}$, ${\dot\Phi}^2$, $\dot { a} \dot {\Phi}$  and the rest to zero as usual, following set of four Noether equations are found as,
\bes\label{2.8}
   6a^2\alpha_{,\Phi} = 0,~~~  \Phi\alpha+a\beta +2a\Phi\alpha_{,a}+a^2\beta_{,a}-\eta a\Phi = 0,\\&
   2\alpha + a \beta_{,\Phi}+a\alpha_{,a}-\eta a= 0,\\&  6k\Phi\alpha+6ak\beta+ 2\kappa( 3\alpha\omega\rho_0a^{-(3\omega+1)}-\eta\rho_0a^{-3\omega} ) -3a^2\alpha U -a^3\beta U_{,\Phi}+ \eta(6ka\Phi - a^3\alpha U)=0.
\end{split}\ee

\subsubsection{Available symmetries in vacuum dominated era.}

In vacuum dominated era the set of solutions of the above equations are the following.\\

\noindent
\textbf{Case-I, $k = \pm 1,0$}. In this case we obtain a pair of solutions.\\

\noindent
\emph{Solution-I}
\be\label{2.2a}
    \alpha= a,~~~ \beta=-2\Phi,~~~\eta=1,~~~ \ U=U_0\Phi^2\ee
\noindent
Now, in view of the definition of $U(\Phi)$ given in \eqref{1.5}, one obtains $U(\Phi) = U_0 \Phi^2 = U_0[F'(R)]^2$, and the quadratic potential produces $F(R)$ and the conserved current in the following forms respectively,
\be\label{2.2a2} F(R) = F_{0} R^2;~~~~~\Sigma = a^3 \dot{R}.\ee

\noindent
\emph{Solution-II}
\be\label{2.2b}
    \alpha={1\over a},~~~ \beta=-\frac{\Phi}{a^2},~~~\eta=0,~~~ \ U= U_0\Phi^3\ee
\noindent
In view of the transformation relations given in \eqref{1.5}, viz. $U(\Phi) = R F'(R)- F(R) =U_0 \Phi^3 = U_0[F'(R)]^3$, one obtains $F(R)$ and the corresponding conserved current as,
\be\label{2.2b2} F(R) = F_{0} R^{\frac{3}{2}};~~~~~\Sigma =  [a \dot\Phi + \dot a \Phi]= \frac{d}{dt}(a\Phi) =\frac{d}{dt}(a\sqrt R).\ee

\noindent
\textbf{Case-II, $k = 0$}. Here again we obtain two sets of solutions, corresponding to different choice of $\eta$.\\

\noindent
\emph{Solution-I}
\be\label{2.2c}\alpha = 0 ,\;\;\;\;\beta = -\beta_{0}a^{-\frac{3}{2}}{\Phi^{-\frac{1}{2}}} ,\;\;\;\;\eta= \frac{\beta_{0}}{2}a^{-\frac{3}{2}}{\Phi^{-\frac{3}{2}}},\;\;\; U(\Phi) = U_{0}{\Phi}^\frac{ 1}{2} , \;\;\;\; {\Sigma} =  \dot a\sqrt{\frac{a}{\Phi}}.\ee
\noindent
Therefore, in view of the transformation relations $U(\Phi) = R F'(R)- F(R) =U_0 \Phi^{\frac{1}{2}} = U_0[F'(R)]^\frac{1}{2}$, one obtains the following form of $F(R)$ and the conserved current,
\be\label{2.2c1} F(R) =\frac{F_{0}}{R};~~~~~{\Sigma} =  R\dot a\sqrt a.\ee

\noindent
\emph{Solution-II}
\be\label{2.2d}\alpha =\alpha_{0} a^{-\frac{1}{2}} ,\;\;\;\;\beta =-\alpha_{0} a{^{-\frac{3}{2}}} \Phi ,\;\;\;\;\eta=\frac{\alpha_{0}}{2} a^{-\frac{3}{2}},\;\;\; U_{0}{\Phi}^\frac{ 7}{2}  , \;\;\;\; {\Sigma} = \sqrt a \dot{a}\Phi+a^\frac{3}{2}\dot{\Phi}.\ee
\noindent
Therefore, in view of the transformation relations $U(\Phi) = R F'(R)- F(R) =U_0 \Phi^{\frac{ 7}{2}} = U_0[F'(R)]^\frac{7}{2}$, one obtains the following forms of $F(R)$  and the conserved current,

\be\label{2.2d1} F(R) = F_{0}R^{\frac{7}{5}};~~~~~{\Sigma} = \sqrt a {d\over dt}(a R^{2\over 5}).\ee

\subsubsection{Available symmetries in radiation dominated era.}
In radiation era $\omega=\frac{1}{3}$, the set of solutions of the Noether equations for $k = 0, \pm 1$ are
\be\label{2.2e}
    \alpha= a,~~~ \beta=-2\Phi, ~~~\eta = 1, ~~~ \ U=U_0\Phi^2\ee
\noindent
Rewriting $U(\Phi)$, as $U(\Phi) = R F'(R)- F(R) =U_0 \Phi^2 = U_0[F'(R)]^2$, the quadratic potential produces the following forms of $F(R)$ and the associated conserved current,
\be\label{2.2e1} F(R) = F_{0} R^2;~~~~~ a^3\dot \Phi = a^3 \dot{R}.\ee

\subsubsection{Available symmetries in pressureless dust era.}
In matter dominated era $\omega=0$, the set of solutions of the Noether equations for $k = \pm 1, 0$ are

\be\label{2.2f}
    \alpha={1\over a},~~~ \beta=-\frac{\Phi}{a^2}, ~~~\eta=0,~~~ \ U= U_0\Phi^3.\ee
Therefore, in view of the transformation relations $U(\Phi) = R F'(R)- F(R) =U_0 \Phi^3 = U_0[F'(R)]^3$, one obtains the following forms of $F(R)$ and its conserved current

\be\label{2.2f1} F(R) = F_{0} R^{\frac{3}{2}};~~~~~\Sigma = [a \dot\Phi + \dot a \Phi]= \frac{d}{dt}(a\Phi) =\frac{d}{dt}(a\sqrt R).\ee
So, in this manner, using the proposed symmetry generator which involves the Hamiltonian constraint\eqref{1.10}, it has also  been possible to recover all the lost symmetries, viz. ($F(R) \propto R^2$, $F(R) \propto R^{-1}$, and $F(R) \propto R^{7\over 5}$) in the Jordan's frame action corresponding to $F(R)$ theory of gravity \eqref{1.1}.

\subsection{Symmetries available in Einstein's frame of reference.}

The fact that symmetries other than $F(R) \propto R^2$ (which appears due to scale invariance of the action and may be explored directly from the field equations), and $F(R) \propto R^{3\over 2}$ (the trivial Noether symmetry obtained following Lagrange multiplier technique and from Jordan's frame of reference \cite{9}) exist, was signalled when the same was attempted to explore in view of analogous Einstein's frame action \eqref{1.1}. Unfortunately, it was not possible to find the trivial symmetry \cite{9} associated with $F(R) \propto R^{3\over 2}$, from the analogous Einstein's frame action. Here we attempt to find the lost symmetry in view of the proposed symmetry generator \eqref{1.10}. The point Lagrangian \eqref{1.7} corresponding to the action (\ref{1.1}) in Robertson-Walker metric \eqref{1.3} leads to the following Hamiltonian in configuration space \cite{1},

\bes\label{2.9} \tilde H_{\tilde t}[\tilde a,\tilde \phi,\dot{\tilde a},\dot{\tilde \phi}]=\frac{3}{\kappa}(-\tilde a \dot{\tilde a}^2+ k \tilde a)+\left({1\over 2}\dot{\tilde\phi}^2 + \tilde V(\tilde\phi)\right)\tilde a^3+\rho_0\tilde a^{-3w}e^{-{\sqrt {\kappa\over 6}}(1-3w)\tilde \phi},\end{split}\ee
\noindent
which is constrained to vanish. Here again, in view of the vector field $X$,

\be\label{2.10} X = \alpha \frac{\partial }{\partial \tilde{a}}+\beta\frac{\partial }{\partial \tilde{\phi}}+\dot\alpha \frac{\partial }{\partial\dot {\tilde{a}}}+ \dot\beta\frac{\partial }{\partial\dot{ \tilde{\phi}} } \ee
\noindent
relative to the Lagrangian $\tilde L_{\tilde t}$ (\ref{1.7}) in the tangent space $[\tilde a,\dot{\tilde a},\tilde\phi,\dot{\tilde \phi}]$, the existence condition for symmetry, $\pounds_X L -\eta H = X L -\eta H = 0 $ \eqref{1.10}, leads to the following differential equation,
\bes\label{2.11}\frac{\alpha}{\kappa} \left[-3\dot{\tilde{a}}^2+3k + 3\kappa\tilde{a}^2\left({1\over 2}\dot{\tilde\phi}^2 - \tilde V(\tilde\phi)\right)+3\kappa\omega \rho_0\tilde a^{-3w}e^{-{\sqrt {\kappa\over 6}}(1-3w)\tilde \phi}\right] \\&+ \beta\left[-\tilde a^3 \tilde V_{,\tilde\phi}+\sqrt {\kappa\over 6}(1-3w)\rho_0\tilde a^{-3w}e^{-{\sqrt {\kappa\over 6}}(1-3w)\tilde \phi}\right]+\left(\alpha_{,\tilde a}\dot {\tilde a}+\alpha_{,\tilde{\phi}}\dot{\tilde \phi}\right)\left[\frac{-6\tilde{a}\dot{\tilde{a}}}{\kappa}\right]+\left(\beta_{,\tilde a}\dot {\tilde a}+\beta_{,\tilde{\phi}}\dot \tilde{ \phi}\right)\left[\tilde{a}^3\dot{\tilde{\phi}}\right]\\&-\eta\left[\frac{3}{\kappa}(-\tilde a \dot{\tilde a}^2+ k \tilde a)+\left({1\over 2}\dot{\tilde\phi}^2 + \tilde V(\tilde\phi)\right)\tilde a^3+\rho_0\tilde a^{-3w}e^{-{\sqrt {\kappa\over 6}}(1-3w)\tilde \phi}\right]   =0.\end{split}\ee
\noindent
Equating coefficients of ${\dot{\tilde a}}^2$, ${\dot{\tilde\phi}}^2$, $\dot {\tilde a} \dot {\tilde\phi}$  and the rest to zero as usual, following set of four Noether equations are found,

\bes\label{2.12} \alpha +6\tilde a \alpha_{,\tilde a}-\eta \tilde a=0,~~~
 \frac{3\alpha}{2} + \tilde a \beta_{,\tilde\phi}-\frac{\eta \tilde a}{2}=0,~~~
 {\tilde a}^2\beta_{,\tilde a}-\frac{6 \alpha_{,\tilde\phi}}{\kappa}=0,\\&
\alpha\left[\frac{3 k}{\kappa}+3\omega \rho_0\tilde a^{-3w}e^{-{\sqrt {\kappa\over 6}}(1-3w)\tilde \phi} -3\tilde a^2\tilde V\right] - \beta\left[\tilde a^3 \tilde V_{,\tilde\phi}- \sqrt {\kappa\over 6}(1-3w)\rho_0\tilde a^{-3w}e^{-{\sqrt {\kappa\over 6}}(1-3w)\tilde \phi}\right]\\&-\eta \left[\frac{3\tilde{a} k}{\kappa}
 +\tilde a^3 \tilde V(\tilde\phi)+\rho_0\tilde a^{-3w}e^{-{\sqrt {\kappa\over 6}}(1-3w)\tilde \phi}\right]=0.\end{split}\ee

\subsubsection{Available symmetries in pure vacuum era.}

\textbf{Case-I, for $k = \pm 1, 0$}.\\

\noindent
In this case, in view of the Noether equations \eqref{2.12}, we obtain following two sets of solutions.\\

\noindent
\emph{Solution-I}\\
The first set of solutions is,
\bes\label{2.3a}
    \alpha = 0,~~~~ \beta = \beta_0, ~~~ \eta =0,~~~ \tilde V(\tilde\phi) = V_{0},
\end{split}\ee
In view of these solutions, the form of $F(R)$ is obtained using the relation \eqref{1.7}, and the expression for the associated conserved current are found as,
\be\label{2.3a1} F(R) =  F_{0}R^{2},\;\;\;\Sigma = \tilde{a}^3\dot{\tilde{\phi}} = a^3\dot R\ee
where, we have used transformation relations \eqref{1.2}, \eqref{1.9} and the form of $F(R)$ so obtained, to translate the conserved current in terms of proper time.\\

\noindent
\emph{Solution-II}\\
The second set of solutions is
\bes\label{2.3a3}
 \alpha = -\frac{\kappa \exp\left({\sqrt{2\kappa\over 3}\tilde \phi}\right)}{6\tilde a},~~~
 \beta = \frac{1}{2\tilde a^2}{\sqrt{2\kappa\over 3}\exp\left({\sqrt{2\kappa\over 3}\tilde \phi}\right)},~~~
 \eta =\frac{\kappa \exp\left({\sqrt{2\kappa\over 3}\tilde \phi}\right)}{6\tilde{a}^2},~~~
 \tilde V(\tilde\phi) = V_{0}\exp\left({\sqrt{2\kappa\over 3}\tilde \phi}\right).
 \end{split}\ee

\noindent
Here the form of $F(R)$ using the relation \eqref{1.7}, and the expression of conserved current are found as,
\be\label{2.3a4} F(R) =  F_{0}R^{\frac{3}{2}};~~~\Sigma =\left(\dot{\tilde a}+\frac{1}{2}\sqrt{2\kappa\over 3}\tilde a\dot{\tilde{\phi}}\right)\exp\left({\sqrt{2\kappa\over 3}\tilde \phi}\right) = \frac{d}{dt}(a\sqrt R)\ee
As mentioned, we have translated the conserved current with respect to the proper time, using the transformation relations \eqref{1.2}, \eqref{1.9} and the above form of $F(R)$. Thus we have retrieved the symmetry which was lost from Einstein's frame action.\\

\noindent
\textbf{Case-II, (For k = 0)}\\

\noindent
Choosing the space to be flat a priori, here again we obtain two sets of solutions.\\

\noindent
\emph{Solution-I}\\
Here the form of $\alpha$, $\beta$, $\eta$ and $\tilde V(\tilde\phi)$ obtained in view of the set of equations \eqref{2.12} are
\bes\label{2.3b}
 \alpha = \sqrt{1\over \tilde a} \exp\left[{-\sqrt{3\kappa\over 8}\tilde \phi}\right],~~~
    \beta = \sqrt{6\over\kappa \tilde a^3} \exp\left[{-\sqrt{3\kappa\over 8}\tilde \phi}\right],~~~
    \eta=0,~~~
    \tilde V(\tilde\phi) =  V_{0}\exp\left[{-\sqrt{3\kappa\over 2}\tilde \phi}\right].
  \end{split}\ee
\noindent
The form of $F(R)$ is found in view of the above form of potential and the expression of the same given in \eqref{1.7}. Form of $F(R)$ and the expression of conserved current found following the same procedure as before, are
\be\label{2.3b1} F(R) =  \frac{ F_{0}}{R},~~~\Sigma = \left[\sqrt{1\over \tilde a}\left(-\frac{6\tilde a \dot{\tilde a}}{\kappa}\right)+\sqrt{6\over\kappa}\tilde a^\frac{3}{2}\dot {\tilde \phi}\right]\exp\left[{-\sqrt{3\kappa\over 8}\tilde \phi}\right] = \sqrt{a} \dot a  R.\ee

\noindent
\emph{Solution-II}\\
The second set of solutions is obtained as,
\bes\label{2.3b4}
 \alpha = \sqrt{1\over \tilde a} \exp\left[{\sqrt{3\kappa\over 8}\tilde \phi}\right],~~~
     \beta = - \sqrt{6\over\kappa \tilde a^3} \exp \left[{\sqrt{3\kappa\over 8}\tilde \phi}\right],~~~
     \eta=0,~~~
     \tilde V(\tilde\phi) =  V_{0}\exp\left[{\sqrt{3\kappa\over 2}\tilde \phi}\right].
   \end{split}\ee

\noindent
The form of $F(R)$ and the expression of conserved current are found following the same procedure as before, in the form
\be\label{2.3b5} F(R) =  F_{0}R^{\frac{7}{5}},\;\;\;\Sigma = \left[\sqrt{1\over \tilde a}\left(-\frac{6\tilde a \dot{\tilde a}}{\kappa}\right)-\sqrt{6\over\kappa}\tilde a^\frac{3}{2}\dot {\tilde \phi}\right]\exp\left[{\sqrt{3\kappa\over 8}\tilde \phi}\right] =  a^{\frac{1}{2}} \dot a  R^{\frac{2}{5}}+ \frac{2}{5}a^{\frac{3}{2}} R^{-\frac{3}{5}} \dot R.\ee

\subsubsection{Available symmetries in radiation dominated era}

In radiation dominated era $\omega=\frac{1}{3}$, for $k = \pm 1, 0$, the forms of $\alpha$, $\beta$, $\eta$ and $\tilde V(\tilde\phi)$ obtained in view of the set of equations \eqref{2.12} as
\bes\label{2.3c}
  \alpha = 0, ~~~ \beta = \beta_0,~~~ \eta =0,~~~ \tilde V(\tilde\phi) = V_{0}
    \end{split}\ee
\noindent
Here the form of $F(R)$ and the expression of conserved current obtained using the same technique as before, are
\be\label{2.3c1} F(R) =  F_{0}R^{2};~~\Sigma = \tilde{a}^3\dot{\tilde{\phi}} = a^3 \dot R.\ee

\subsubsection{Available symmetries in pressure-less dust era.}

In the matter dominated era $\omega=0$, for arbitrary choice of curvature parameter, $k = \pm 1, 0$ the forms of $\alpha$, $\beta$, $\eta$ and $\tilde V(\tilde\phi)$ obtained from the set of equations \eqref{2.12} are
 \bes\label{2.3d}
 \alpha = -\frac{\kappa \exp\left({\sqrt{2\kappa\over 3}\tilde \phi}\right)}{6\tilde a},~~~
     \beta = \frac{1}{2\tilde a^2}{\sqrt{2\kappa\over 3}\exp\left({\sqrt{2\kappa\over 3}\tilde \phi}\right)},~~~
     \eta =\frac{\kappa \exp\left({\sqrt{2\kappa\over 3}\tilde \phi}\right)}{6\tilde{a}^2},~~~
     \tilde V(\tilde\phi) = V_{0}\exp\left({\sqrt{2\kappa\over 3}\tilde \phi}\right).
     \end{split}\ee

\noindent
The form of $F(R)$ and the expression for the conserved current are
\be\label{2.3d1} F(R) =  F_{0}R^{\frac{3}{2}},\;\;\;\Sigma = \left(\dot{\tilde a}+\frac{1}{2}\sqrt{2\kappa\over 3}\tilde a\dot{\tilde{\phi}}\right)\exp\left({\sqrt{2\kappa\over 3}\tilde \phi}\right)=\frac{d}{dt}(a\sqrt R).\ee

Hence, we have been able to recover the lost symmetry corresponding to $F(R) \propto R^{3\over 2}$, which appears both in vacuum and matter dominated era in view of the proposed symmetry generator \eqref{1.10}. Interestingly, in the process we have not lost any symmetry that were found earlier.

\section{Concluding remarks}

We always felt that matching classical field equations is not enough to establish true mathematical equivalence between different canonical prescriptions of the same theory, since canonical formulation of higher order theory is non-trivial. It is also required that all the different canonical actions produce all the symmetries available in the background of a specific metric. In an earlier manuscript \cite{1}, we observed that Jordan's and Einstein's frames produce different symmetries, which couldn't be translated back and forth using the transformation relation, which applies in matching classical field equations. In view of a modified existence condition for Noether symmetry  $\pounds_X L - \eta H$, which includes the Hamiltonian constraint, it has been possible to extract all the existing symmetries in all the three canonical actions corresponding to $F(R)$ theory of gravity, independently. Since, the existence condition is general (not restricted to a minisuperspace) it is possible to establish equivalence in different anisotropic metrics also which results from Bianchi classification. Moreover, in the inhomogeneous situations, where momenta constraint ($p_i = 0, i = 1, 2, 3$) exist, it is required to test if the symmetry generator $\pounds_X L - \eta H - \sum {\delta_i p_i} = 0$ applies equally. These issues will be posed in future.\\

While preparing the manuscript, through a private communication from Dr. Nikolaos Dimakis, we learnt that Christodoulakis et al \cite{ch1} had already suggested that in order to explore all the symmetries of a gravitational Lagrangian, one must not fix the gauge ($N$), the lapse function a priori. In the process more symmetries were explored \cite{ch2}. This is because, in that case, the ($^0_0$) equation of Einstein ( essentially the Hamilton constraint equation in configuration space variables) which appears from ${\partial L\over \partial N} = 0$ is lost. This means one has to find the symmetries keeping the lapse $N$ arbitrary. In that case the generator \eqref{2.2} in Lagrange multiplier technique, should be modified to

\be\label{3.1} X = \alpha \frac{\partial }{\partial {a}}+\beta\frac{\partial }{\partial R}+\dot\alpha \frac{\partial }{\partial\dot {a}}+\dot\beta \frac{\partial }{\partial \dot{R}} + \gamma \frac{\partial }{\partial {N}} +\dot\gamma \frac{\partial }{\partial\dot {N}},\ee
that in Jordan's frame \eqref{2.6} should be replaced by

\be\label{3.2} X = \alpha \frac{\partial }{\partial {a}}+\beta\frac{\partial }{\partial \Phi}+\dot\alpha \frac{\partial }{\partial\dot {a}}+\dot\beta \frac{\partial }{\partial \dot{\Phi}} + \gamma \frac{\partial }{\partial {N}} +\dot\gamma \frac{\partial }{\partial\dot {N}},\ee
and finally in Einstein's frame \eqref{2.10} should be replaced by,

\be\label{3.2} X = \alpha \frac{\partial }{\partial \tilde{a}}+\beta\frac{\partial }{\partial \tilde{\phi}}+\dot\alpha \frac{\partial }{\partial\dot {\tilde{a}}}+ \dot\beta\frac{\partial }{\partial\dot{ \tilde{\phi}} }+ \gamma \frac{\partial }{\partial {N}} +\dot\gamma \frac{\partial }{\partial\dot {N}}.\ee
In fact, this is what we have already observed. In order to translate field equations from one frame to the other, the constraint equation is required. On the contrary, Noether symmetry does not recognize the constraint. This is the reason why we have suggested to enter the constraint equation through a Lagrange multiplier into the existence condition of symmetry. Hence, this treatment is identical to the one suggested by Christodoulakis et al \cite{ch1}. The only problem is involving the lapse the Lagrangian becomes singular \cite{ch3}. Nevertheless, this may be handled following Dirac's constraint analysis.\\

\noindent
\textbf{Acknowledgement:}\\
We would like to thank Dr. Nikolaos Dimakis, for bringing our attention to three recent papers, cited at the end, and explaining in brief the formalism adopted there.

\end{document}